\documentclass[preprint,prb,showpacs]{revtex4-2}

\usepackage{graphicx}
\usepackage{dcolumn}
\usepackage{bm}
\usepackage{natbib}
\usepackage[english]{babel}
\usepackage{amsmath}
\usepackage{float}

\begin{document}

\title{\boldmath Hybridization gap and $f$-electron effect evolutions with Cd- and Sn-doping in CeCoIn$_5$ via infrared spectroscopy \unboldmath}

\author{Myounghoon Lee} \author{Yu-Seong Seo} \author{Seulki Roh} \author{Seokbae Lee} \author{Jihyun Kim} \author{Tuson Park} \author{Jungseek Hwang}\email{jungseek@skku.edu}
\affiliation{Department of Physics, Sungkyunkwan University, Suwon, Gyeonggi-do 16419, Republic of Korea}

\date{\today}

\begin{abstract}

We investigated hole (Cd)- and electron (Sn)-doped CeCoIn$_5$ (CeCo(In$_{1-x}T_x$)$_5$ ($T$ = Cd or Sn)) using infrared spectroscopy. Doping-dependent hybridization gap distribution functions were obtained from the optical conductivity spectra based on the periodic Anderson model formalism. The hybridization gap distribution exhibits two components: in-plane and out-of-plane hybridization gaps. The doping-dependent evolution of the two gaps indicated that the out-of-plane gap was more sensitive to doping. Furthermore, the magnetic optical resistivity exhibited a doping-dependent evolution of the $f$-electron amplitude. The two dopant types exhibited different physical properties depending on the level of doping. The Sn dopant increases the $f$-electron amplitude, whereas the Cd dopant does not affect the $f$-electron amplitude. Doping-dependent effective mass is peaked at pure (or undoped) CeCoIn$_5$. Our spectroscopic results may help understand the doping-dependent electronic evolution of one of the canonical heavy fermion systems, CeCoIn$_5$.

\end{abstract}


\maketitle

\section{Introduction}

Heavy fermion systems are one type of intermetallic compound containing elements with 4$f$ or 5$f$ electrons in unfilled electron bands that exhibit intriguing and unique electronic and magnetic properties. Heavy fermion systems exhibit a Sommerfeld coefficient $\gamma$ up to 1000 times larger than that expected from the free electron model \cite{steglich:1979}. The large Sommerfeld coefficient results from a large effective mass owing to the strong band renormalization near the Fermi level, in which $f$-electrons are involved. Heavy fermion systems also exhibit other interesting phenomena, such as unconventional superconductivity, magnetic ordering, and an insulating state. Therefore, many studies on this topic have been conducted \cite{anderson:1959,steglich:1979,millis:1987,donovan:1997,shishido:2002,christianson:2004,harrison:2004,park:2006,silhanek:2006,willers:2010,stockert:2011,gofryk:2012,thompson:2012,seo:2013,chen:2016,wirth:2016}. Several optical studies have been conducted on the heavy fermion systems \cite{bonn:1988,degiorgi:1999,singley:2001,dordevic:2001,dressel:2002,hancock:2004,mena:2005,burch:2007,nagel:2012,lee:2023}. Optical studies have observed and discussed the gradual development of the hybridization gap with decreasing temperature and the increased effective mass of charge carriers at low temperatures. Furthermore, a theoretically predicted universal relationship between the hybridization gap and the coherent temperature was observed \cite{dordevic:2001}. The hybridization gap distribution functions of Ce$M$In$_5$ ($M$ = Co, Rh, and Ir) compounds have been obtained and reported \cite{burch:2007}. The hybridization gap distribution function of CeCoIn$_5$ has been obtained using the maximum entropy method (MEM) and the temperature-dependent $f$-electron amplitude has been obtained from the measured optical resistivity spectra \cite{lee:2023}.

CeCoIn$_5$ is known to be located near the quantum critical point (QCP) in the Doniach phase diagram \cite{haga:2001,coleman:2005,shishido:2005,settai:2007,hu:2012}. CeCoIn$_5$ exhibits a peculiar temperature-dependent resistivity with some characteristic temperatures, such as the coherent temperature ($T^*$) and Kondo temperature ($T_K$). Numerous experiments have demonstrated that CeCoIn$_5$ is at the QCP by controlling external parameters, such as doping, pressure, and magnetic field \cite{petrovic:2001,paglione:2003}. When the In atom in CeCoIn$_5$ was replaced by either Cd or Sn, the system can be moved from the QCP in the phase diagram. Unlike the full replacement of Co with either Rh or Ir, Cd or Sn doping on the In site required only a relatively small percentage of dopants to similarly affect on the ground state (or QCP). Because of the small amount of doping, the structural changes caused by In-site doping were insignificant \cite{chen:2018,bauer:2006}. Therefore, to understand the nature of QCP, many studies on Cd- and Sn-doped CeCoIn$_5$ have been conducted \cite{daniel:2005,bauer:2005,bauer:2006,pham:2006,donath:2006,nicklas:2007,booth:2009,seo:2013,sakai:2015,chen:2018,chen:2019a}.
Cd-doped CeCoIn$_5$ was hole-doped, whereas Sn-doped CeCoIn$_5$ was electron-doped. In addition to the doping type, there were some intriguing differences between the two dopants. A nuclear quadrupole resonance study \cite{sakai:2015} showed that Cd doping introduces a heterogeneous electron state and decreases the local $f$-$p$ hybridization, whereas Sn doping introduces a homogeneous electron state and increases the global hybridization strength. An X-ray absorption study showed that the spatial configuration of the 4$f$ wave function could be a good probe for studying small changes in the hybridization of 4$f$ and conduction electrons \cite{chen:2018}. There were one in-plane In atom (In(1)) and four out-of-plane In atoms (In(2)) in the unit cell. Therefore, if In atoms in the two (in-plane and out-of-plane) In sites are randomly replaced by dopant atoms (Cd or Sn), the in-plane In(1) will be replaced by 20\% of dopants and the out-of-plane In(2) will be replaced by 80\% of dopants. However, previous studies with an extended X-ray absorption fine-structure (EXAFS) showed that the in-plane In(1) was preferentially replaced by the dopants: 43\% and 55\% of Cd and Sn, respectively \cite{daniel:2005,booth:2009}. Cd doping induces a long-range magnetic order, and the Cd dopants in Cd-doped CeCoIn$_5$ can be used as a tuning agent between superconductivity and antiferromagnetism. The applied pressure globally suppresses the Cd-induced magnetic order and restores superconductivity \cite{pham:2006}. Cd dopants can act as nuclei for spin droplets in Cd-doped CeCoIn$_5$ under high pressure \cite{seo:2013}. The local and global effects of Cd and Sn dopants, respectively, are attractive issues in heavy fermion physics. Optical studies on Cd- and Sn-doped CeCoIn$_5$ samples have not yet been conducted. An optical study on Cd- and Sn-doped CeCoIn$_5$ may provide interesting new information for understanding the nature of the QCP of CeCoIn$_5$. 

In this study, we investigated Cd- and Sn-doped CeCoIn$_5$, CeCo (In$_{1-x}T_x$)$_5$ ($T$ = Cd or Sn), using infrared spectroscopy. The optical conductivity spectra at various doping levels at various temperatures were obtained from the measured reflectance spectra in a wide spectral range (80 - 25000 cm$^{-1}$) using the Kramers-Kronig analysis. From the optical conductivity, we observed that the hybridization gap evolved with doping; the gap increased as Sn doping increases, whereas it decreased as Cd-doping increased. The hybridization gap distribution functions of Cd- and Sn-doped CeCoIn$_5$ were obtained using a model-independent maximum entropy method (MEM) based on the periodic Anderson model. The resulting gap distribution function was properly fitted with only two Gaussian peaks; the gap distribution function consists of two (small and large) components. The small and large components were assigned as in-plane and out-of-plane gaps, respectively \cite{lee:2023}. The doping-dependent properties of the two hybridization gaps were obtained from the doping-dependent properties of the magnitudes, positions, and widths of the two Gaussian peaks. The out-of-plane gap exhibited a significant doping dependence. In addition, by examining the magnetic optical resistivity \cite{lee:2023}, we found that the 4$f$ amplitude did not change as Cd doping increased, but increased with Sn doping. Furthermore, the doping-dependent effective mass of charge carriers was obtained using the extended Drude model. The effective mass of pure CeCoIn$_5$ showed a maximum value, which is consistent with previous experimental results \cite{coleman:2005,coleman:2007} showing that the effective mass diverges as the material system approaches to the QCP. Our optical results provide doping- and temperature-dependent evolutions of the hybridization gap and $f$-electron amplitude in the CeCo(In$_{1-x}T_x$)$_5$ ($T$ = Cd or Sn) systems. We expect that our results may help us to understand the doping-dependent electronic evolution of CeCoIn$_5$.

\section{Experiment}

The doping- and temperature-dependent optical properties of the Cd- and Sn-doped CeCoIn$_5$ were determined using infrared spectroscopy. The single crystal samples (1.2 and 2.0\% Cd-doped (hole-doped) CeCoIn$_5$ and 1.2, 2.0, and 3.6\% of Sn-doped (electron-doped) CeCoIn$_5$) were grown using the In self-flex method. It is worth noting that the doping concentration in this study is the actual one. Two different (nominal and actual) definitions of doping concentrations have been used \cite{pham:2006,nicklas:2007,chen:2018}. In general, the difference between the two doping concentrations is significant; for example, a $\sim$10\% nominal Cd doping concentration ends up being only 1\% actual Cd doping one \cite{nicklas:2007}. A detailed description of the growth method can be found in the literature \cite{petrovic:2001,macaluso:2002,hu:2013}. The 1.2\% and 2.0\% Cd-doped samples are denoted as Cd 1.2\% and Cd 2.0\%, respectively. Similarly, the 1.2, 2.0, and 3.6 \% Sn-doped samples are denoted as Sn 1.2\%, Sn 2.0\% and Sn 3.6\%, respectively. The DC resistivity data of all the samples as functions of temperature were taken by using a four-probe technique. The DC resistivity data can be found in the Supplementary Materials (Figs. S1(a) and (b)). The reflectance spectra of each sample in a wide spectral range (80 - 25000 cm$^{-1}$) at various temperatures were measured using a commercial Fourier transform infrared spectrometer (Bruker Vertex 80v) and a liquid helium flow cryostat (ARS LT3). An {\it in-situ} metallization method was used to obtain an accurate reflectance spectrum \cite{homes:1993}. The Kramers-Kronig analysis was used to obtain optical constants, including the optical conductivity, from the measured reflectance spectrum \cite{wooten,tanner:2019}. To perform the Kramers-Kronig integration, the measured spectrum in a finite spectral range must be extrapolated to both zero and infinity. For the extrapolation to zero, the Hagen-Rubens relation ($1-R(\omega) \propto \sqrt{\omega}$) was used. Here, the Hagen-Rubens relation was determined by using the measured DC resistivity data. For the extrapolation to infinity, $R(\omega) \propto \omega^{-2}$ was used from 25000 to 10$^6$ cm$^{-1}$ and, above 10$^6$ cm$^{-1}$, the free electron behavior ($R(\omega) \propto \omega^{-4}$) was assumed.

\section{Results and discussion}

\begin{figure}[!htbp]
  \vspace*{-0.1 cm}%
 \centerline{\includegraphics[width=5.5 in]{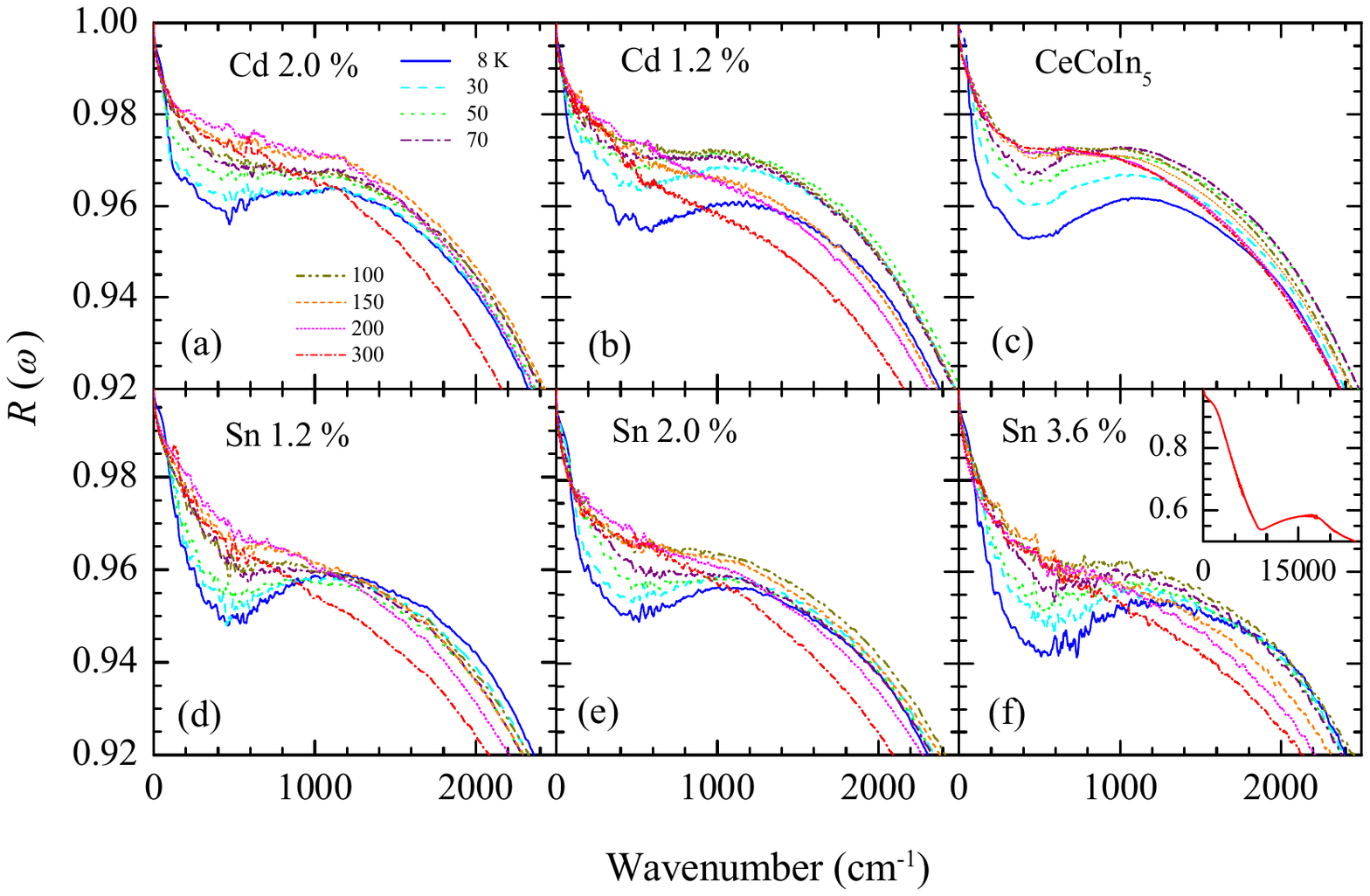}}%
  \vspace*{-0.3 cm}%
\caption{(Color online) Measured reflectance spectra of all six samples, including undoped CeCoIn$_5$, at various temperatures. In the inset of the frame (f), the measured reflectance spectrum of the 3.6\% Sn-doped sample at 300 K is shown in a wide spectral range up to 25000 cm$^{-1}$.}
\label{fig1}
\end{figure}

\begin{figure}[!htbp]
  \vspace*{-0.1 cm}%
 \centerline{\includegraphics[width=5.5 in]{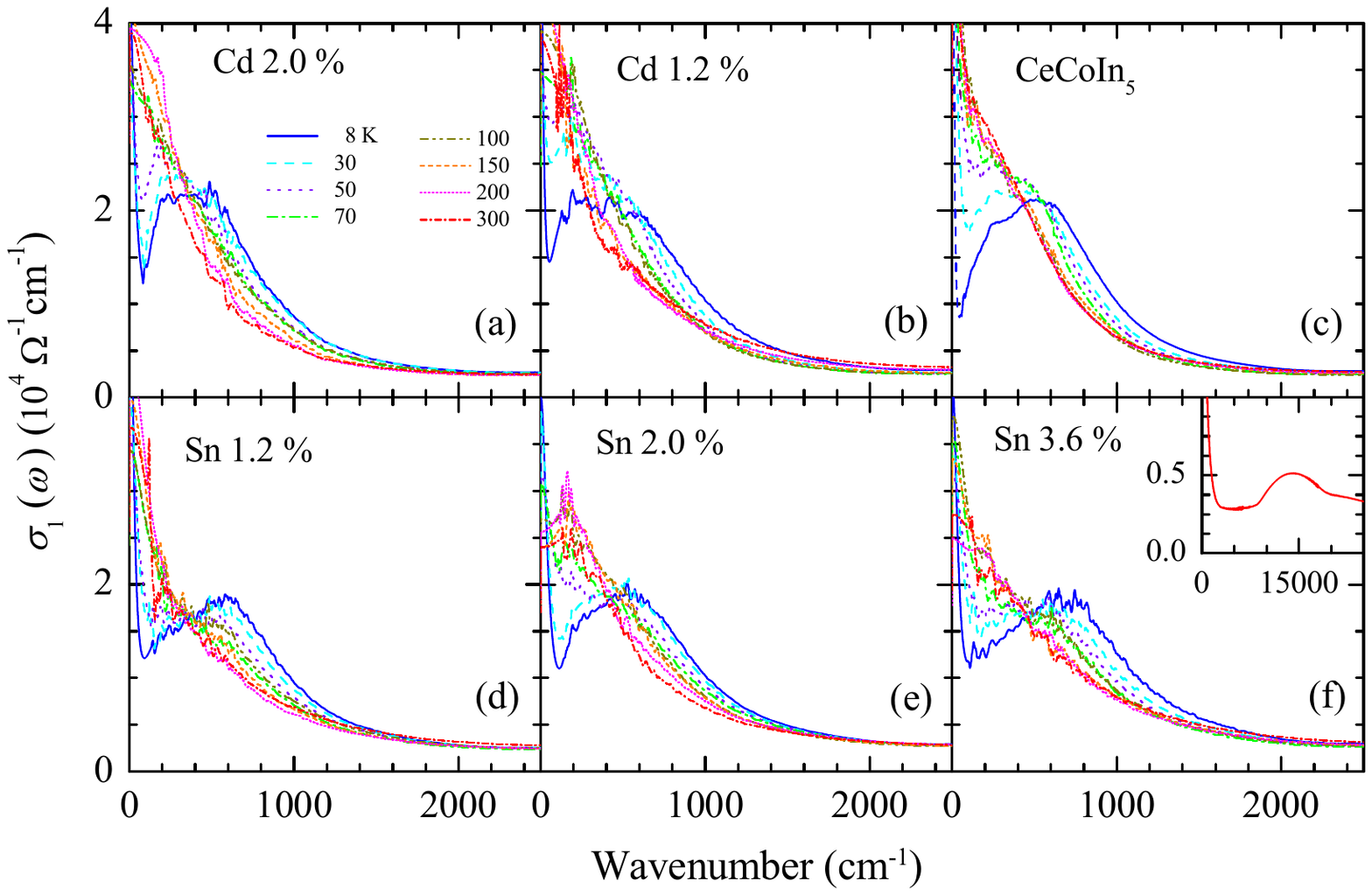}}%
  \vspace*{-0.3 cm}%
\caption{(Color online) Optical conductivity spectra of all six samples at various temperatures, obtained from the measured optical reflectance using the Kramers-Kronig analysis. In the inset of the frame (f), the optical conductivity of the 3.6\%-doped sample at 300 K is shown in a wide spectral range up to 25000 cm$^{-1}$.}
\label{fig2}
\end{figure}

\begin{figure}[!htbp]
  \vspace*{-0.1 cm}%
 \centerline{\includegraphics[width=5.5 in]{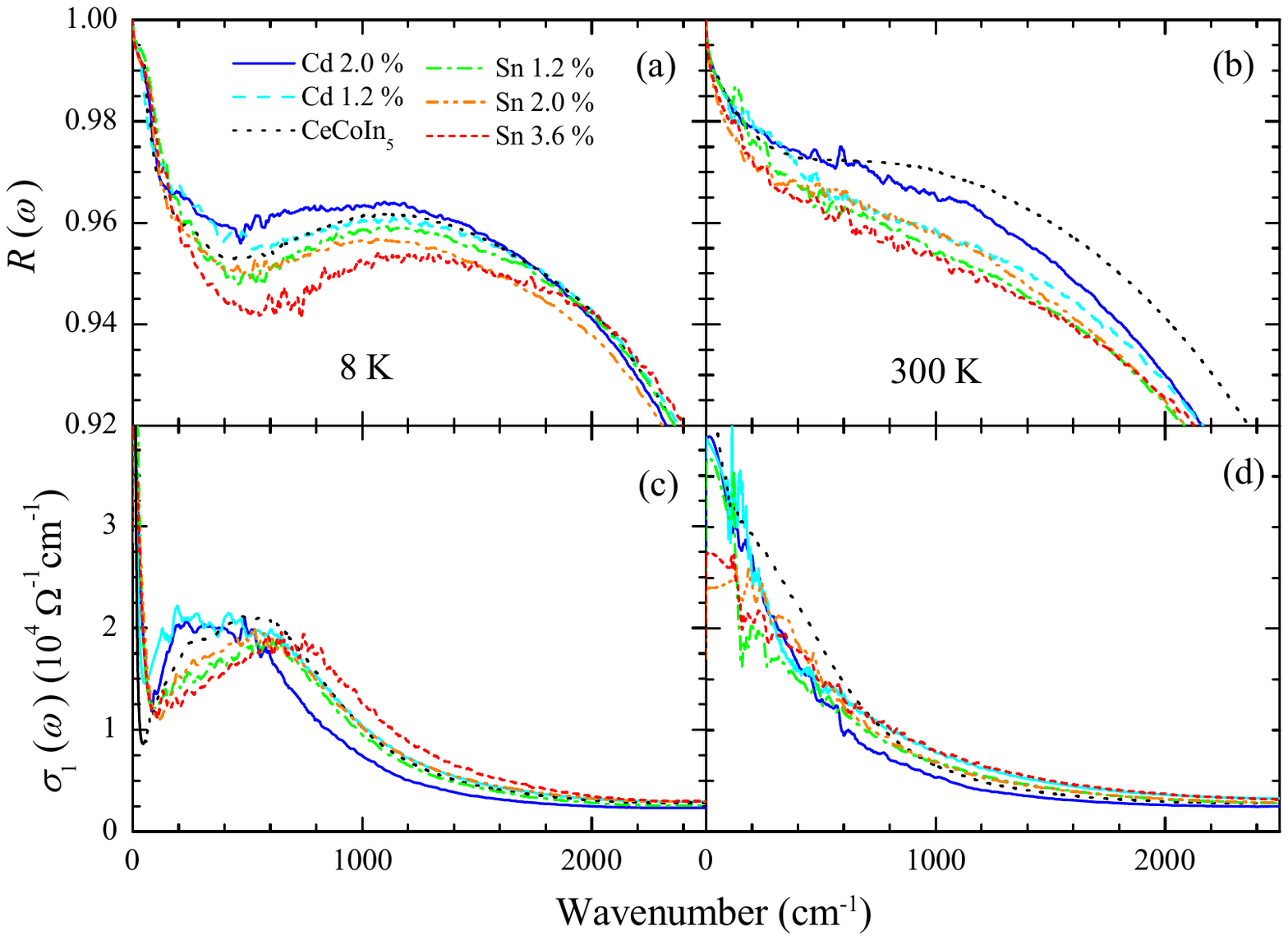}}%
  \vspace*{-0.3 cm}%
\caption{(Color online) Reflectance and optical conductivity spectra of all six samples at 8 and 300 K, for better comparisons of doping dependence.}
\label{fig3}
\end{figure}

Fig. \ref{fig1} shows the measured reflectance spectra of all six samples, including pure CeCoIn$_5$, at various temperatures. All six samples exhibited similar overall temperature- and frequency-dependent trend. At 300 K, the reflectance spectra were almost featureless (see Fig. \ref{fig3}(b)). However, as the temperature drops, a dip near 500 cm$^{-1}$ appears and is getting deeper. The dip can be a characteristic feature of an optical gap in the reflectance spectrum, resulting from the spectral weight transfer from low to high energy (see Fig. \ref{fig2}). However, below the dip, the reflectance rapidly increases and approaches 1.0, resulting in a metallic ground state, which can be described by the Drude model with a small scattering rate. The inset of Fig. \ref{fig1}(f) shows the measured reflectance spectrum of the 3.6\% Sn-doped sample at 300 K in a wide spectral range up to 25000 cm$^{-1}$. The reflectance at high energies showed minimal temperature dependence. However, the reflectance at low energy and low temperature changes significantly; at a given temperature (8 K), the dip becomes deeper as the doping increases from hole (Cd) to electron (Sn), as shown in Fig. \ref{fig3}(a). Therefore, we focused on the measured spectra at low frequencies to discuss doping- and temperature-dependent evolution of the electronic structures.

The optical conductivity spectra of all six samples obtained from the measured reflectance spectra using the Kramers-Kronig analysis are shown in Fig. \ref{fig2}. The optical conductivity spectra of all samples showed similar temperature- and frequency-dependent trends. At 300 K, the optical conductivity spectra of all the samples appeared similar, with a broad Drude mode at low frequencies (see Fig. \ref{fig3}(d)). As the temperature drops, a broad peak near 300 cm$^{-1}$ appears, grows, and shifts to higher energy, along with a very sharp Drude-like mode at the low-energy side. In fact, the broad peak is known to have two components, which have been observed previously \cite{lee:2023} and assigned to the in-plane and out-of-plane hybridization gaps based on the LDA+DMFT calculations \cite{shim:2007}. The broad peak also shows significant doping dependence, as shown in Fig. \ref{fig3}(c), which displays the optical conductivity spectra of all six samples at 8 K. As the doping changed from hole to electron, the peak shifted to higher energies, indicating that the size of the hybridization gap increased.

Fig. \ref{fig3} shows the reflectance and optical conductivity spectra of all six samples at 8 and 300 K. Fig. \ref{fig3}(a) shows the reflectance spectra of all six samples at 8 K. Interestingly, the doping-dependent trend of reflectance is similar to the temperature-dependent trend of reflectance, as shown in Fig. \ref{fig1}. The effect of the increase in doping from hole to electron is similar to that of the temperature decrease. As the doping changes from hole- to electron-doped, the dip near 500 cm$^{-1}$ gets deeper and shifts to higher energy. The dip position did not change significantly with the temperature (see in Fig. 1). Fig. \ref{fig3}(b) shows the reflectance spectra of all six samples at 300 K. The reflectance shows doping dependance; as the doping changes from holes to electrons, the reflectance decreases almost monotonically. Corresponding features can be observed in the optical conductivity spectra shown in Figs. \ref{fig3}(c) and (d).

\begin{figure}[!htbp]
  \vspace*{-0.1 cm}%
 \centerline{\includegraphics[width=5.5 in]{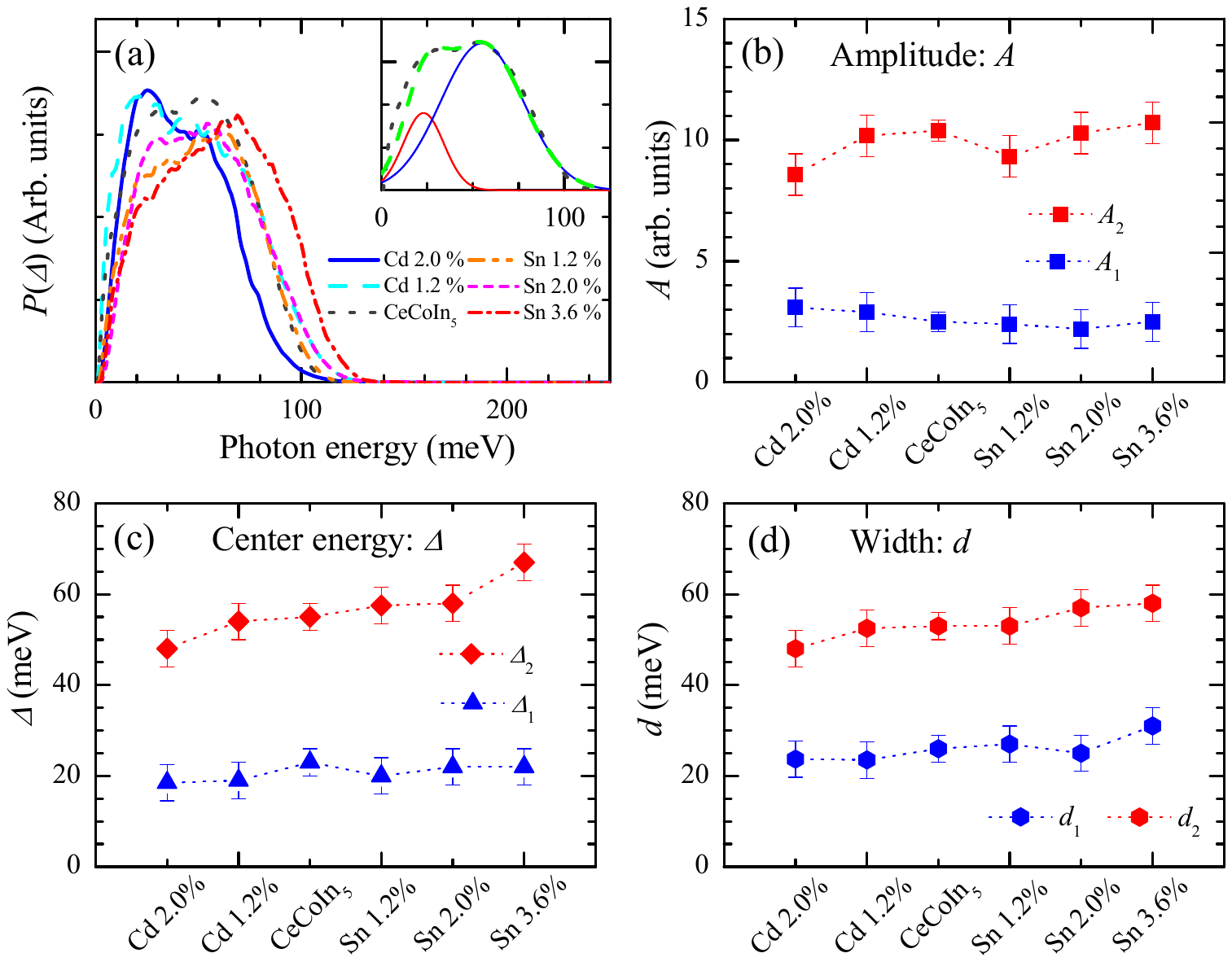}}%
  \vspace*{-0.3 cm}%
\caption{(Color online) Doping-dependent hybridization gap distribution function. (a) Hybridization gap distribution functions of all six samples at 8 K. In the inset, the gap distribution function of CeCoIn$_5$ is shown along with a fit using two Gaussian peaks. The amplitudes (b), center energies (c), and widths (d) of the two Gaussian peaks as functions of doping are displayed.}
\label{fig4}
\end{figure}

To more quantitatively analyze the doping-dependent evolution of the hybridization gap ($\Delta$), the hybridization gap distribution functions ($P(\Delta)$) were obtained using a method introduced by previous studies \cite{burch:2007,lee:2023}. Note that because the hybridization gap is $k$ dependent, the hybridization gap in a measured optical spectrum may appear as a gap distribution function ($P(\Delta)$) \cite{burch:2007}. In this method, the incoherent part of the optical conductivity ($\sigma_1^{\mathrm{incoh}}(\omega)$) at 8 K is described in terms of the periodic Anderson model ($\sigma_1^{\mathrm{PAM}}(\omega, \Delta)$) as $\sigma_1^{\mathrm{incoh}}(\omega) = \int_0^{\omega_c}P(\Delta)\sigma_1^{\mathrm{PAM}}(\omega, \Delta)d\Delta$, where $\omega_c$ is a cutoff frequency. Note that $\sigma_1^{\mathrm{incoh}}(\omega)$ is the optical conductivity subtracted by the sharp Drude mode. $P(\Delta)$ was obtained from the measured optical conductivity by solving the inversion problem using the MEM with the known kernel, $\sigma_1^{\mathrm{PAM}}(\omega, \Delta) = A\Theta(\omega-\Delta)/\sqrt{\omega^2-\Delta^2}$, where $A$ is a constant and $\Theta(z)$ is the Heaviside step function \cite{lee:2023}. The MEM allows us to obtain the most probable result from the measured data. The incoherent optical conductivity data and MEM fits at 8 K are shown in Fig. S2 in the Supplementary Materials. The obtained gap distribution functions ($P(\Delta)$) of all six samples at 8 K are shown in Fig. \ref{fig4}(a). The gap distribution function monotonically shifts to high energy as the doping changes from hole (Cd) to electron (Sn).

The obtained gap distribution function mainly consists of two components that can be fitted with two Gaussian peaks, as shown in the inset of Fig. \ref{fig4}(a). The two components are assigned to the in-plane and out-of-plane hybridization gaps \cite{shim:2007,lee:2023}; the component located at lower (higher) energy corresponds to the in-plane (out-of-plane) gap. The amplitude ($A$), center energy ($\Delta$), and width ($d$) of each Gaussian peak are determined from the fitting parameters of the two-Gaussian fit. All the fitting parameters are shown in Figs. \ref{fig4}(b-d). The amplitudes, center energies, and widths of the two hybridization gaps exhibited different doping dependencies. The amplitude ($A_1$) of the in-plane gap slightly decreased, whereas the amplitude ($A_2$) of the out-of-plane gap slightly increased as the doping changes from hole to electron, indicating that the overlap between the In(2) 5$p$ orbital in the out-of-plane direction and the Ce 4$f$ orbital may have increased. As the doping changes from hole to electron doping, the in-plane gap ($\Delta_1$) slightly increases, while the out-of-plane gap ($\Delta_2$) significantly increases. The size ($\Delta$) of the hybridization gap is proportional to the coupling constant $J$ between the local moment and conduction electrons. As the doping changed from hole to electron, the out-of-plane hybridization strength increased whereas the in-plane strength did not change significantly. Our results show that the out-of-plane hybridization gap is more sensitive to doping than the in-plane gap, even though the dopants preferentially occupy the in-plane In(1) site \cite{daniel:2005}. Interestingly, both widths ($d_1$ and $d_2$) increased as the doping changes from hole to electron, for which we do not yet clearly know the reason.

\begin{figure}[!htbp]
  \vspace*{-0.1 cm}%
 \centerline{\includegraphics[width=5.5 in]{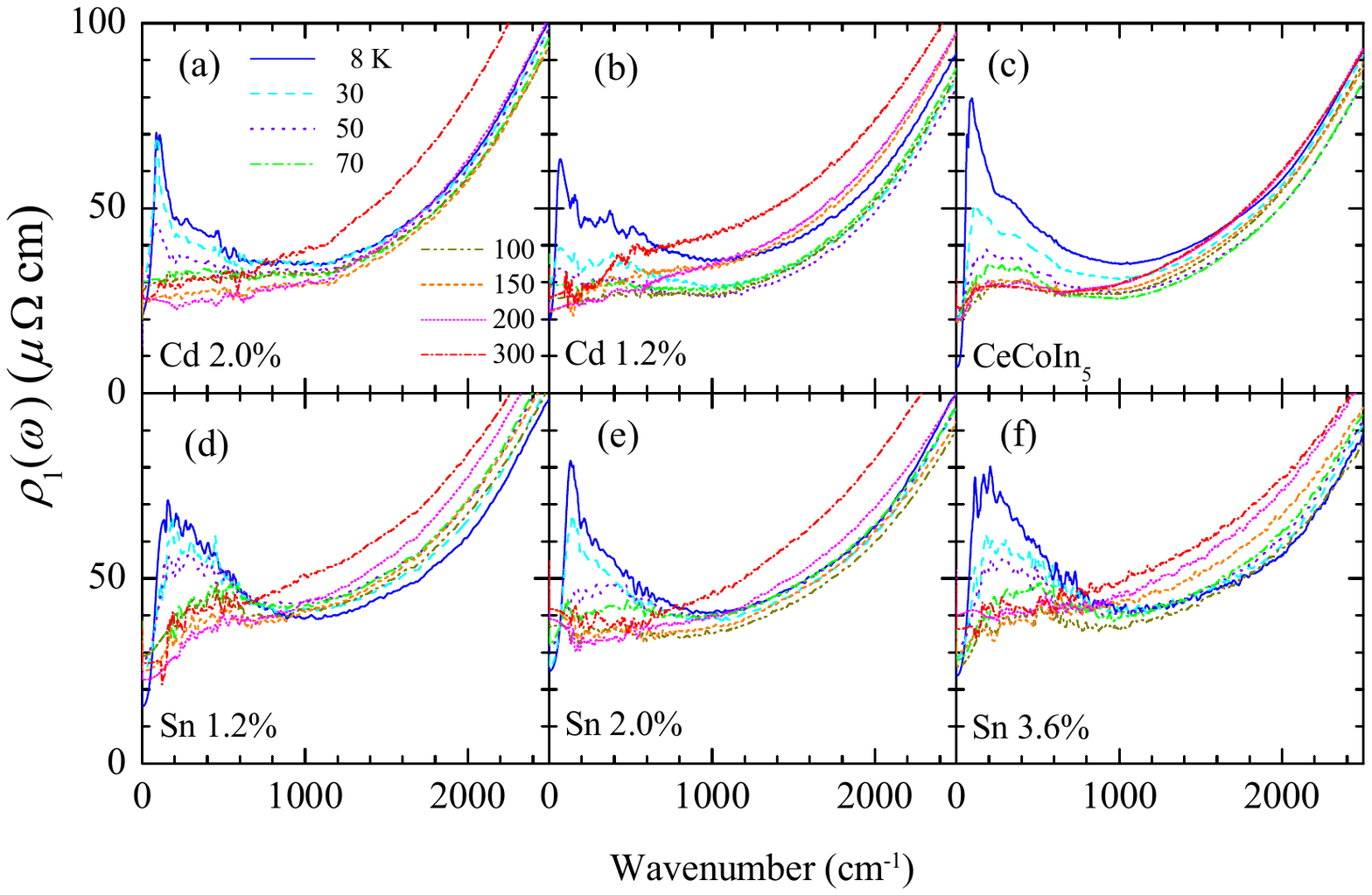}}%
  \vspace*{-0.3 cm}%
\caption{(Color online) Optical resistivity spectra of all six samples at various selected temperatures.}
\label{fig5}
\end{figure}

The real parts of the optical resistivity of all six samples, including the pure CeCoIn$_5$, are shown in Fig. \ref{fig5}. The complex optical resistivity ($\tilde{\rho}(\omega)$) is defined as $\tilde{\rho}(\omega) \equiv 1/\tilde{\sigma}(\omega)$, where $\tilde{\sigma}(\omega)$ is the complex optical conductivity \cite{nagel:2012}. Each doped sample exhibited similar temperature- and frequency-dependent behavior to the pure sample. At 300 K, there is almost no peak at low energy, but as the temperature decreases, a broad peak near 200 - 400 cm$^{-1}$ appears, grows, and shifts to lower energy. The optical (or frequency-dependent) resistivity spectra at 8 K resemble the temperature-dependent DC resistivity data (see Figs. S1(a) and (b)). The peak was identified as a coherent peak in the frequency domain in a previous optical study \cite{lee:2023}. We observed a doping-dependent coherent peak in the optical resistivity (see Fig. 6(b)).

\begin{figure}[!htbp]
  \vspace*{-0.1 cm}%
 \centerline{\includegraphics[width=5.5 in]{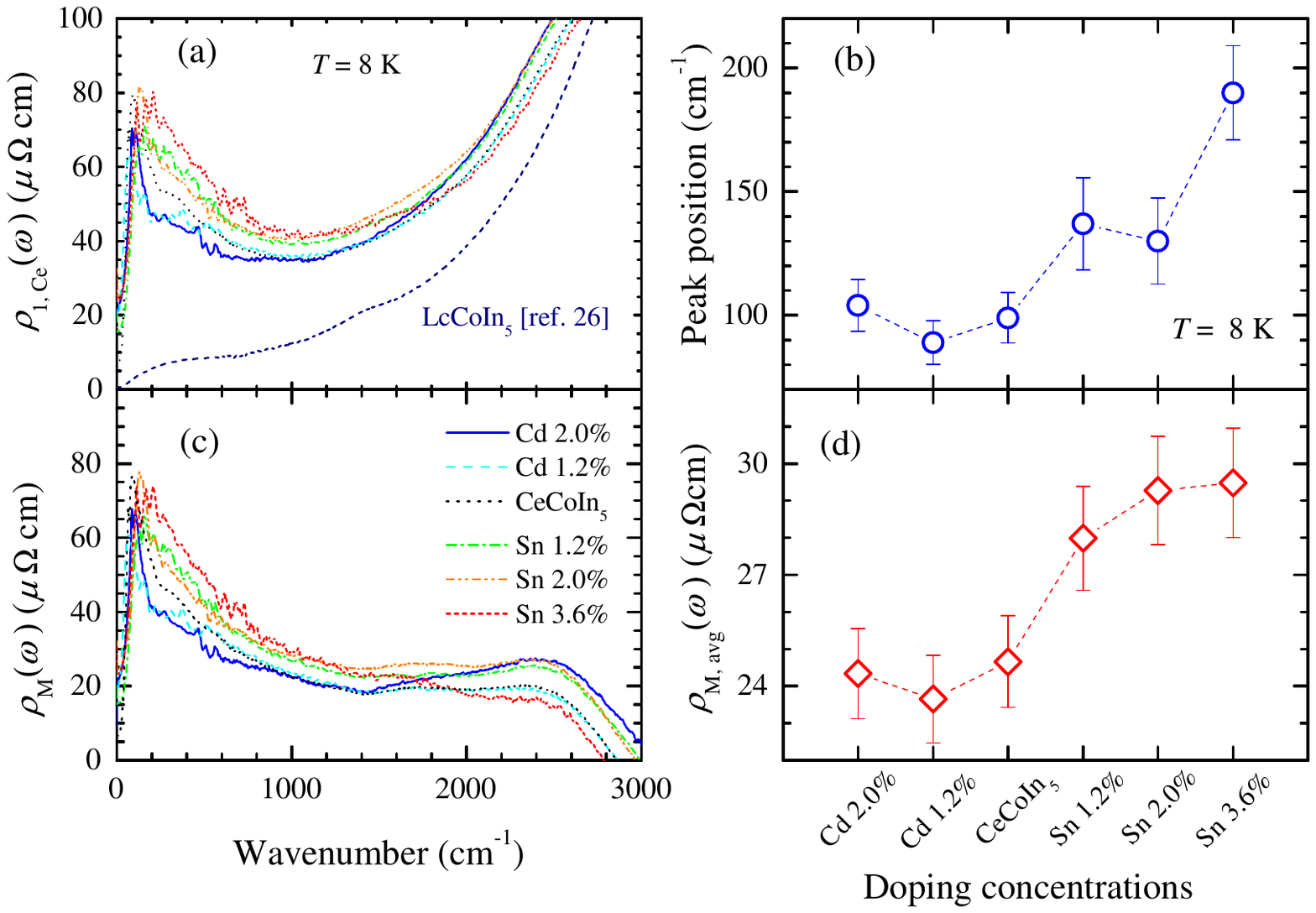}}%
  \vspace*{-0.3 cm}%
\caption{(Color online) (a) Optical resistivity spectra of all six samples and LaCoIn$_5$ at 8 K. (b) Doping-dependent peak energy in the optical resistivity at 8 K. (c) Magnetic optical resistivity spectra of all six samples at 8 K. (d) Doping-dependent average magnetic optical resistivity at 8 K. }
\label{fig6}
\end{figure}

To observe the doping-dependent properties at low temperatures more clearly, the optical resistivity spectra of all six samples at 8 K are shown in Fig. \ref{fig6}(a). The coherence peak in the frequency domain (or optical coherent peak) exhibits a peculiar doping-dependent behavior as shown in Fig. \ref{fig6}(b); as the doping changes from hole to electron, the peak energy is almost constant in the hole-doping region but increases significantly as electron doping increases. We evaluated the peak position as follows: We fitted the peak region with a quartic function of frequency (see Fig. S3 in the Supplementary Materials) and obtained the frequency at the maximum of the fitting curve. The width of the coherent peak roughly increased as the doping increases from holes to electrons. It is worth noting that the optical coherent peak and the coherent temperature ($T^*$) exhibit different doping-dependent trends; the $T^*$ increases almost linearly from hole to electron doping (see Fig. S1 in the Supplementary Materials), whereas the optical coherent peak shows a different doping-dependent trend. Therefore, these two features may not be closely related to each other. The magnetic optical resistivity ($\rho_M(\omega)$) was obtained by subtracting the optical resistivity of LaCoIn$_5$ ($\rho_{1,La}(\omega)$) from that of each doped sample ($\rho_1(\omega)$) as in the previous study \cite{lee:2023}, i.e., $\rho_M(\omega) \equiv \rho_1(\omega)-\rho_{1, La}(\omega)$. Note that we used a reported $\rho_{1,La}(\omega)$ spectrum in literature \cite{lee:2023} which is shown in Fig. 6(a). Through this subtraction, the electronic background could be removed; the magnetic contribution from the Ce 4$f$ electrons in the samples will remain in the resulting spectrum, $\rho_M(\omega)$. The resulting magnetic optical resistivity spectra of all six samples at various temperatures are shown in the Supplementary Materials (Fig. S4). Fig. \ref{fig6}(c) shows the resulting magnetic optical resistivity spectra of all six samples at 8 K. It is worth noting that because the optical resistivity is roughly the inverse of the optical conductivity, the two-gap features appear as dips instead of peaks in the optical resistivity spectrum at 8 K, as can be seen in Fig. \ref{fig6}(a).
 
In a previous comparative study of CeCoIn$_5$ and LaCoIn$_5$ [26], the enhanced resistivity of CeCoIn$_5$ compared with that of LaCoIn$_5$ was associated with the 4$f$ electrons in CeCoIn$_5$ and was used to investigate the temperature-dependent evolution of the 4$f$-electron amplitude. Herein, we investigated the doping dependence of magnetic optical resistivity at 8 K. The average magnetic optical resistivity could be defined as $\rho_{M}^{avg} \equiv (1/\omega_c)\int_0^{\omega_c} \rho_M(\omega')d\omega'$, where $\omega_c$ is a cutoff frequency, which depends on the sample \cite{lee:2023}. Note that, for the calculation of the average magnetic optical resistivity, only the positive $\rho_M(\omega)$ is taken into account. The average magnetic optical resistivity of all six samples at various temperatures are shown in the Supplementary Materials (Fig. S5). Fig. \ref{fig6}(d) shows the average magnetic optical resistivity data for all six samples at 8 K. The average magnetic optical resistivity has been known to be intimately associated with the $f$-electron amplitude \cite{lee:2023}. The average magnetic optical resistivity shows a similar doping-dependent behavior to the coherent peak energy in the optical resistivity, as can be seen in Figs. \ref{fig6}(b) and (d); it is also almost constant as the hole doping increases, whereas it appreciably increases as the electron doping increases. The similarity in the doping dependence between the peak energy and average magnetic optical resistivity may indicate that the two quantities are closely related to each other. The dopant-dependent differences in both the peak energy and average magnetic optical resistivity may be associated with the previously observed dopant-dependent difference \cite{sakai:2015}, and the small or negligible changes in both the average magnetic optical resistivity and peak energy with hole doping may be associated with the local effect of Cd (hole) doping. The homogeneous electronic state induced by electron (Sn) doping may result in a significant increase in both the peak energy and $f$-electron amplitude.

\begin{figure}[!htbp]
  \vspace*{-0.1 cm}%
 \centerline{\includegraphics[width=5.5 in]{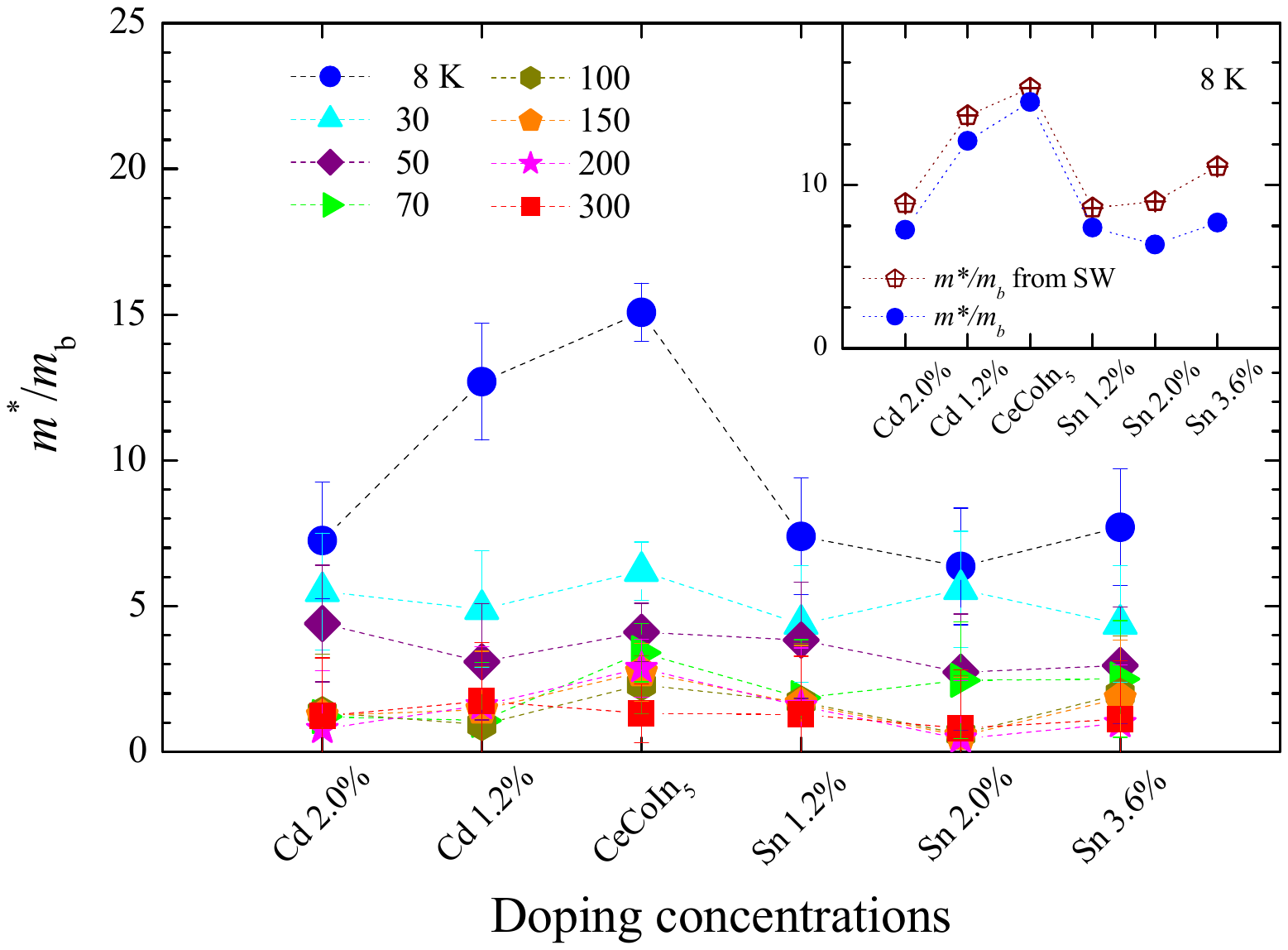}}%
  \vspace*{-0.3 cm}%
\caption{(Color online) Effective masses of all six samples at various temperatures. In the inset, the effective masses obtained using two different methods at 8 K are shown (see the text for detailed descriptions).}
\label{fig7}
\end{figure}

The effective mass of charge carriers in each sample was obtained using the extended Drude model formalism \cite{gotze:1972,allen:1977,puchkov:1996,hwang:2004}. In the extended Drude model, the optical effective mass ($m^*(\omega)/m_b$) can be written as $m^*(\omega)/m_b \equiv -(\Omega_p^2/4 \pi \omega) \Im{[1/\tilde{\sigma}(\omega)]}$, where $\Omega_p$ is the plasma frequency of the charge carriers, and $m^*(\omega)$ and $m_b$ are the optical effective mass and the band mass, respectively. Note that the plasma frequency ($\Omega_p$) is obtained from the optical spectral weight up to 2500 cm$^{-1}$, i.e., $\Omega_p^2 = (120/\pi) \int_0^{2500 \:\mbox{cm}^{-1}}\sigma_1(\omega)d\omega$, where all frequencies are in cm$^{-1}$ units. The optical effective mass spectra of all six samples at various temperatures are shown in the Supplementary Materials (Fig. S6). Fig. \ref{fig7} shows the effective mass ($m^*(0, T)/m_b$) of all six samples. By lowering the temperature, all samples showed an increase in the effective mass resulting from the hybridization gap formation. The effective mass of pure CeCoIn$_5$ was the largest among all the samples. This is closely related to the QCP of CeCoIn$_5$. In principle, at the QCP, an infinite effective mass is expected at zero temperature \cite{stewart:2001,coleman:2005} owing to abundant quantum fluctuations. Therefore, our results confirm that CeCoIn$_5$ is located near the QCP. In the inset, we display the effective masses of all six samples at 8 K obtained using a different method, based on the spectral weight, i.e., $m^*/m_b = \int_0^{2500 \:\mbox{cm}^{-1}} \sigma_1(\omega, 300 \:\mbox{K})d\omega/\int_0^{\omega_d}\sigma_1(\omega, 8\:\mbox{K})d\omega$, where $\omega_d$ is the frequency at the dip in the optical conductivity below the hybridization gap \cite{singley:2001,lee:2023}. The results obtained using two different methods agree well with each other. Note that, at 8 K, the dip frequency ($\omega_d$) divides the spectral weight into two components: coherent and incoherent components. The spectral weight above $\omega_d$ is the incoherent component. Therefore, the mass enhancement of the quasiparticles is associated only with the coherent component below $\omega_d$ \cite{singley:2001}. In this case, the long tail of the narrow Drude curve above $\omega_d$ looks omitted. However, almost the same amount of the omitted spectral weight from the tail of the hybridization gap below $\omega_d$ is added. Therefore, the overall spectral weight below $\omega_d$ is more or less equal to the narrow Drude spectral weight. Furthermore, our measured lowest data point is rather high because of the small size of the doped samples. The high cutoff frequency may cause some uncertainty in the estimated effective mass. However, when we used the measured DC resistivity data for the extrapolation from the lowest data point to zero in the Kramers-Kronig process, the estimated $m^*/m_b$'s were quite reliable (see Fig. S7 in the Supplementary Materials).  

\begin{figure}[!htbp]
  \vspace*{-0.1 cm}%
 \centerline{\includegraphics[width=5.5 in]{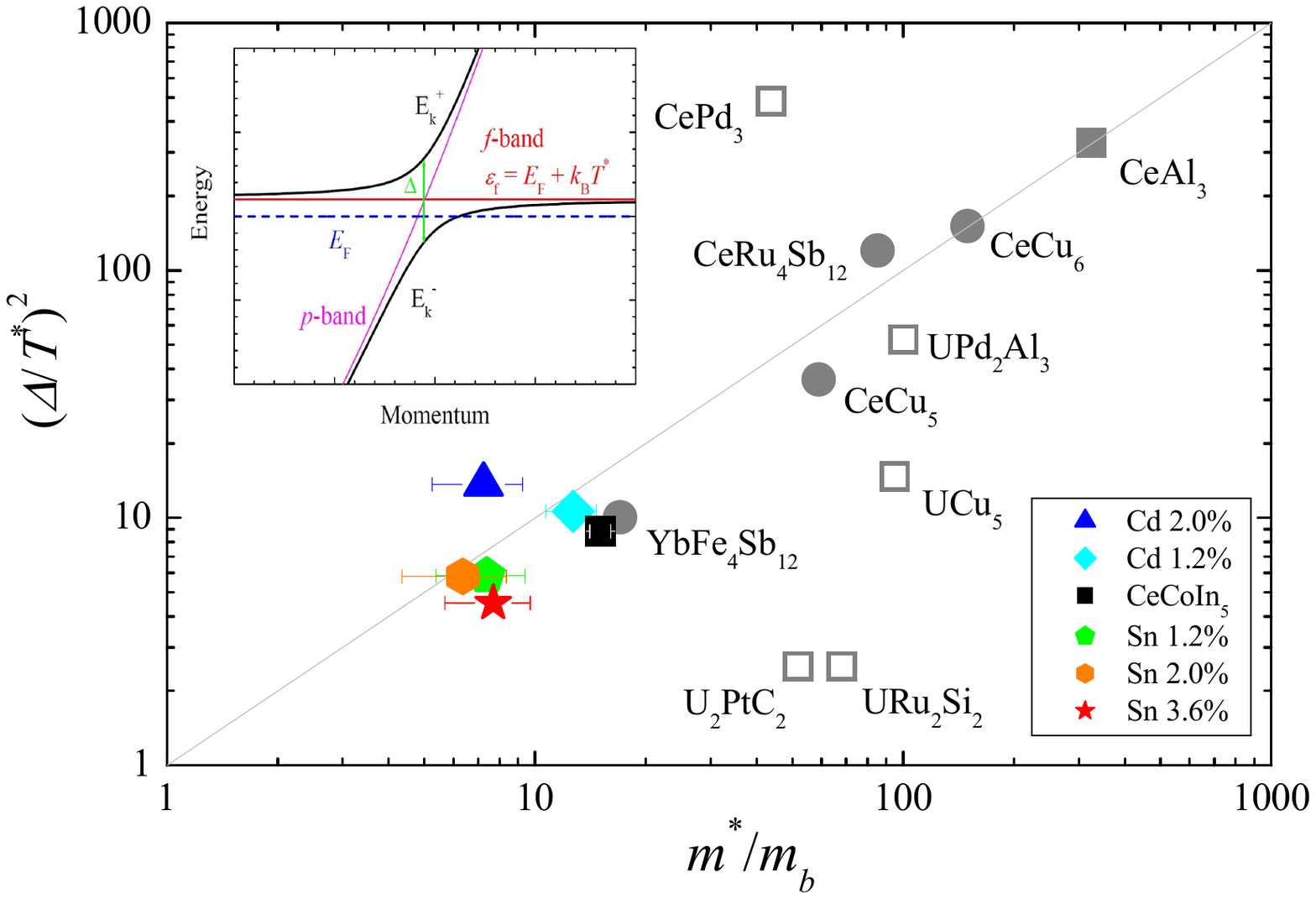}}%
  \vspace*{-0.3 cm}%
\caption{(Color online) Universal relation between the effective mass $m^*/m_b$ and the magnitude of the hybridization gap $(\Delta/T^*)^2$ of nonmagnetic heavy fermion compounds. The colored symbols are added by this study. In the inset, a schematic diagram of the renormalized electronic band structure near the Fermi level by the band hybridization formation is shown. Here $E_k^-$ and $E_k^+$ are the upper and lower bands, respectively, resulting from the hybridization of the parabolic $p$- and flat $f$-bands and $\Delta$ is the hybridization gap. $E_f$ and $T^*$ are the Fermi energy and coherence temperature, respectively. $k_B$ is the Boltzmann constant.}
\label{fig8}
\end{figure}

A theoretical study introduced a universal relation between the effective mass and the hybridization gap, i.e., $(\Delta/T^*)^2 \cong m^*/m_b$ \cite{millis:1987,millis:1987a} based on the energy-momentum dispersion of the hybridization gap (see the inset of Fig. \ref{fig8}), where $\Delta$ is the hybridization gap, $T^*$ is the coherent temperature, and $m^*/m_b$ is the effective mass with respect to the band mass, $m_b$. This universal relation is expected in paramagnetic heavy fermion systems \cite{dordevic:2001}. An experimental study observed a universal relation from the measured spectra of various nonmagnetic heavy fermion systems, as shown in Fig. \ref{fig8} \cite{dordevic:2001}. Note that the uranium-based heavy fermion compounds are consistently off from the universal relation. In the compounds, the magnetic excitations might further add the effective mass to the value from the hybridization \cite{dordevic:2001}. We added data points (colored symbols) to the universal line plot. Note that the coherent temperatures ($T^*$) of our samples were obtained from the measured DC resistivity (Fig. S1(c)) and the hybridization gaps are the smaller in-plane hybridization gaps ($\Delta_1$) as in the previous study \cite{dordevic:2001}. The Cd- and Sn-doped CeCoIn$_5$, including pure CeCoIn$_5$, were reasonably well fitted by the universal line. 

\section{Conclusions}

We investigated the evolution of optical (or electronic) properties by replacing the In atoms in CeCoIn$_5$ with a small percentage of Sn and Cd dopants. From this study, we determined the doping-dependent hybridization gap and $f$-electron amplitude. The hybridization gap distribution function $P(\Delta)$ was obtained by using the model-independent MEM based on the periodic Anderson model. The obtained gap distribution function consisted of two Gaussian peaks, which were identified as the in-plane and out-of-plane hybridization gaps. Based on these results, we investigated the doping-dependent properties of the two (in-plane and out-of-plane) hybridization gaps. We found that as the doping changes from hole to electron, the hybridization strength between Ce 4$f$ orbital and out-of-plane In(2) 5$p$ orbital increased significantly, although in-plane In(1) was preferentially replaced by the dopants. This observation indicates that the out-of-plane hybridization is more sensitive to doping. Furthermore, we confirmed that Cd and Sn doping affected optical (electronic) properties differently. The doping-dependent behavior of the coherent peak energy in the optical resistivity is similar to that of the average magnetic optical resistivity, indicating that two quantities are closely associated with each other. The average magnetic optical resistivity is closely related to the $f$-electron amplitude. We expect that our findings will be helpful to understand the electronic evolution of CeCoIn$_5$ with temperature and the doping.

%
%
\acknowledgments JH acknowledges the financial support from the National Research Foundation of Korea (NRFK Grant Nos. 2020R1A4A4078780 and 2021R1A2C101109811). This research was supported by BrainLink program funded by the Ministry of Science and ICT through the National Research Foundation of Korea (2022H1D3A3A01077468).

\bibliographystyle{apsrev4-2}
\bibliography{bib}

\end{document}